\documentclass[letter,traditabstract]{aa}

\usepackage{graphics}
\usepackage{graphicx}
\usepackage{multicol}
\usepackage{color,soul}
\usepackage{txfonts}
\usepackage{amsmath}
\usepackage{verbatim}
\usepackage{pdflscape}
\usepackage{afterpage}

\usepackage{natbib,twoopt}
\usepackage{amsmath} 
\usepackage{hyperref} 
\bibpunct{(}{)}{;}{a}{}{,} 
\makeatletter
\newcommandtwoopt{\citeads}[3][][]{\href{http://adsabs.harvard.edu/abs/#3}%
{\def\hyper@linkstart##1##2{}%
\let\hyper@linkend\@empty\citealp[#1][#2]{#3}}}

\newcommandtwoopt{\citepads}[3][][]{\href{http://adsabs.harvard.edu/abs/#3}%
{\def\hyper@linkstart##1##2{}%
\let\hyper@linkend\@empty\citep[#1][#2]{#3}}}

\newcommandtwoopt{\citetads}[3][][]{\href{http://adsabs.harvard.edu/abs/#3}%
{\def\hyper@linkstart##1##2{}%
\let\hyper@linkend\@empty\citet[#1][#2]{#3}}}

\newcommandtwoopt{\citeyearads}[3][][]%
{\href{http://adsabs.harvard.edu/abs/#3}
{\def\hyper@linkstart##1##2{}%
\let\hyper@linkend\@empty\citeyear[#1][#2]{#3}}}

\newcommandtwoopt{\citeturl}[3][][]{\href{#3}%
{\def\hyper@linkstart##1##2{}%
\let\hyper@linkend\@empty\citet[#1][#2]{#3}}}

\newcommandtwoopt{\citepurl}[3][][]{\href{#3}%
{\def\hyper@linkstart##1##2{}%
\let\hyper@linkend\@empty\citep[#1][#2]{#3}}}

\makeatother

\hypersetup{colorlinks=true,linkcolor=blue,citecolor=blue,urlcolor=blue}

\def\inclP{152}
\def\inclPerr{14}
\def\OmegaPval{135}
\def\OmegaPvalerr{34}
\def\correlvalP{-0.15}

\def\inclR{28}
\def\inclRerr{14}
\def\OmegaRval{295}
\def\OmegaRvalerr{34}
\def\correlvalR{-0.22}

\def\Massc{12}
\def\Masscerrplus{+12}
\def\Masscerrminus{-5}

\begin{document}

\title{Orbital inclination and mass of the exoplanet candidate Proxima c}
\titlerunning{Orbital inclination and mass of Proxima c}
\authorrunning{P. Kervella, F. Arenou \& J. Schneider}
%
\author{
Pierre~Kervella\inst{1}
\and
Fr\'ed\'eric Arenou\inst{2}
\and
Jean Schneider\inst{3}
}
\institute{
LESIA, Observatoire de Paris, Universit\'e PSL, CNRS, Sorbonne Universit\'e, Univ. Paris Diderot, Sorbonne Paris Cit\'e, 5 place Jules Janssen, 92195 Meudon, France, \email{pierre.kervella@obspm.fr}.
\and
GEPI, Observatoire de Paris, Universit\'e PSL, CNRS, 5 Place Jules Janssen, 92195 Meudon, France.
\and
LUTH, Observatoire de Paris, Universit\'e PSL, 5 Place Jules Janssen, 92195 Meudon, France
}
\date{Received ; Accepted}
\abstract {We analyze the orbital parameters of the recently discovered exoplanet candidate Proxima~c using a combination of its spectroscopic orbital parameters and Gaia DR2 astrometric proper motion anomaly. We obtain an orbital inclination of $i = \inclP \pm \inclPerr \deg$ for the prograde solution, corresponding to a planet mass of $m_c = \Massc^{\Masscerrplus}_{\Masscerrminus}\,M_\oplus$, comparable to Uranus and Neptune. While the derived orbital parameters are too uncertain to accurately predict the position of the planet for a given epoch, we present a map of its probability of presence relative to its parent star in the coming years.
}
\maketitle

\section{Introduction}

\object{Proxima Centauri} (\object{GJ 551}, \object{HIP 70890}, hereafter Proxima) is a red dwarf of spectral type M5.5V, and our nearest stellar neighbor. It is a member of the $\alpha$\,Centauri triple system (\object{WDS J14396-6050AB}, \object{GJ559AB}), which also comprises the solar-like stars $\alpha$\,Cen A (\object{HD 128620}) and B (\object{HD 128621})  of spectral types G2V and K1V \citepads{2017A&A...597A.137K, 2017A&A...598L...7K},  respectively.
Using the radial velocity technique, \citetads{2016Natur.536..437A} discovered a terrestrial-mass planet orbiting Proxima in its habitable zone (\object{Proxima b}).
\citetads{2020SciA....6.7467D} confirmed its parameters and identified a second planet candidate, \object{Proxima c}, orbiting at 1.5\,au with a minimum mass $m_c \sin i = 6\,M_{\oplus}$.
One of the interests of the planetary system of Proxima is that, because of its proximity to us, it is a privileged target for future interstellar probes \citepads{2017AJ....154..115H, 2018MNRAS.474.3212F}, such  as for example the Breakthrough Starshot project \citepads{2018AcAau.152..370P}.
Here, we combine the spectroscopic orbital parameters of Proxima determined by \citetads{2020SciA....6.7467D} with the astrometric proper motion anomaly (PMa) measured by \citetads{2019A&A...623A..72K}. Using these two complementary observables, we constrain the orbital parameters of the planet, and in particular the orbital plane inclination $i$ and the longitude of the ascending node $\Omega$.

\section{Analysis from spectroscopy and astrometry}

\subsection{Observational quantities\label{observ}}

\begin{table*}[htp]
\caption{Observational parameters of Proxima Centauri and Proxima c.}
\centering
\begin{tabular}{lccl}
\hline \hline
\noalign{\smallskip}
Quantity & & Value & Ref. \\
\hline
\noalign{\smallskip}
Mass of Proxima & $m_\star$ & $0.1221 \pm 0.0022\,M_\odot$ & M15 \\
\noalign{\smallskip}
Parallax & $\varpi$ & $768.529 \pm 0.220$\,mas & GDR2 \\
\noalign{\smallskip}
RV amplitude of Proxima & $K_\mathrm{c}$ & $1.2 \pm 0.4$\,m\,s$^{-1}$ & D20 \\
\noalign{\smallskip}
Orbital period & $P_\mathrm{orb}$ & $1900^{+96}_{-82}$\,days & D20  \\
\noalign{\smallskip}
Inferior conjunction BJD & $T_\mathrm{c,conj}$ & $2\,455\,892^{+101}_{-102}$ & D20 \\
\noalign{\smallskip}
Eccentricity & $e$ & 0 & Fixed \\
\noalign{\smallskip}
GDR2 PM anomaly & $\Delta \mu_\mathrm{G2}$ & $\Delta\mu_\alpha = +0.218 \pm 0.112$\,mas\,a$^{-1}$ & K19 \\
        & & $\Delta\mu_\delta = +0.384 \pm 0.215$\,mas\,a$^{-1}$ & K19 \\
        & & $\rho(\Delta\mu_\alpha, \Delta\mu_\delta) = 0.37$ & GDR2 \\
\noalign{\smallskip}
\hline
\end{tabular}
\tablebib{
D20: \citetads{2020SciA....6.7467D};
GDR2: \citetads{2018A&A...616A...1G};
K19: \citetads{2019A&A...623A..72K};
M15: \citetads{2015ApJ...804...64M}.
}
\label{Proxima-table}
\end{table*}

The spectroscopic orbital parameters summarized in Table~\ref{Proxima-table} were determined by \citetads{2020SciA....6.7467D} based on high-precision radial velocity measurements collected using the HARPS and UVES spectrographs.
These parameters characterize the orbital reflex motion induced by Proxima c on its parent star along the line of sight.

\citetads{2019A&A...623A..72K} define the PMa as the difference between the short-term proper motion (PM) vector from the Hipparcos \citepads[Hip2,][]{2007ASSL..350.....V} or Gaia DR2 \citepads[GDR2,][]{2018A&A...616A...1G} catalogs and the long-term PM vector.
The latter is computed using the difference between the Hip2 and GDR2 positions, taking advantage of the long time baseline of 24.25 years to reach high accuracy.
Historically, this long-term to short-term PM comparison has been employed by \citetads{1844MNRAS...6R.136B} to discover \object{Sirius B} and \object{Procyon B}, and recent applications of this technique can be found in \citetads{2007A&A...464..377F}, \citetads{2008ApJ...687..566M}, \citetads{2018ApJS..239...31B, 2019ApJS..241...39B} and \citetads{2019A&A...623A.116K}.
It relies on the fact that the presence of an orbiting stellar or planetary companion shifts the barycenter of the system away from its photocenter (located very close to the primary star center in the case of a planet).
This results in a deviation of the short-term PM vector (attached to the photocenter) compared to the long-term PM vector (that mostly follows the barycenter motion).
In this paper, we assume the long-term Hip2-GDR2 PM to be the motion of the barycenter of the  Proxima
system (including the star and its planets).
The orbital periods of Proxima~b and c, namely 11.2 days and 5.2 years, respectively, are much shorter than the 24.25 years separating the Hip2 and GDR2 measurements, and their effect on the long-term PM can be neglected.
The influence of the inner planet Proxima~b on the GDR2 PMa vector $\vec{\Delta \mu_\mathrm{G2}}$ is also negligible due to its very short orbital period (11.2\,days) compared to the GDR2 observing window (668\,days).
The PMa vector listed in Table~\ref{Proxima-table} therefore closely traces the tangential reflex motion of Proxima induced by the outer planet Proxima c, averaged over the GDR2 time window.
Further details on the sensitivity function and limitations of the PMa as an indicator of binarity can be found in \citetads{2019A&A...623A..72K}.

Following \citetads{2017A&A...598L...7K}, the mass of \object{Proxima} is estimated to $m_\star = 0.1221 \pm 0.0022\,M_\odot$ from the mass--luminosity relation calibrated by \citetads{2015ApJ...804...64M} and the 2MASS magnitude $m_K = 4.384 \pm 0.033$  \citepads{2006AJ....131.1163S}.
As in \citetads{2019A&A...623A..72K}, we slightly corrected the parallax of Proxima from the Gaia DR2 catalog \citepads{2018A&A...616A...1G} by adding a parallax zero point offset of $+29\,\mu$as (negligible compared to the uncertainty) and   rescaling the error bar, as recommended by \citetads{2018A&A...616A...2L}. We obtain $\varpi = 768.529 \pm 0.220$\,mas for epoch J2015.5, whose uncertainty ($\pm 0.03\%$) is negligible for the present analysis.

\subsection{Orbital parameters and mass of Proxima c \label{orbitparams}} 

We fit the orbital parameters of Proxima~c taking into account the spectroscopic orbital parameters determined by \citetads{2020SciA....6.7467D}, as well as the $\vec{\Delta \mu_\mathrm{G2}}$ PMa vector from \citetads{2019A&A...623A..72K}.
We retrieved the transit times of Proxima on the Gaia detectors from the online Gaia Observation Forecast Tool (GOST)\footnote{\url{https://gaia.esac.esa.int/gost/index.jsp}}.
This allowed us to model the time smearing in the GDR2 catalog PMa using the true distribution of individual measurement epochs corresponding to the PM vector reported in the GDR2 catalog.
We use this information to match the PMa from our orbit model to the measured (averaged) PMa vector.
The effective GDR2 PMa measurement epoch for Proxima is found to be J2015.67.
While this method is in theory more accurate than a straight, unweighted integration over the GDR2 measurement window, we find that both computations agree very well in practice.
This is due to the high density of the Gaia transits and their regular distribution over the observing time window, which covers approximately half of the orbital period of Proxima c.

Similarly to \citetads{2020SciA....6.7467D}, we assume a circular orbit for planet c ($e=0$). The orbital period and the adopted mass of Proxima (Table~\ref{Proxima-table}) define the orbital radius $a_c$. The only orbital parameters to be determined are therefore the orbital inclination $i$ and the longitude of the ascending node $\Omega$.
For the estimation of the uncertainties on $i$ and $\Omega$, we followed a classical Monte Carlo (MC) numerical approach.
We adopted a prior on the orbital inclination proportional to $\sin(i)$ using rejection sampling, which corresponds to a random orientation of the orbit. The choice of this prior is justified by the fact we have a low signal-to-noise ratio ($<5$) on the astrometry and radial velocity data; further details can be found in \citetads{2001A&A...372..935P} and \citetads{2004IAUS..202...60A} for example.
We neglected the uncertainties on the mass of Proxima and its parallax.
We took into account the uncertainties on the spectroscopic orbital parameters, the averaging of the PMa over the GDR2 transit epochs, the PMa vector uncertainty, and the correlation listed in the GDR2 catalog between the PM vector components ($\rho=0.37$).
Due to the fact that we have only one PMa vector, two inclinations are possible: $0^\circ \leqslant i_1 \leqslant 90^\circ$ (retrograde) and $i_2 = 180^\circ - i_1$ (prograde, $90 \leqslant i_2 \leqslant 180^\circ$).
Following the standard convention, $\Omega$ is counted from north ($\Omega=0^\circ$) toward east, and corresponds to the position angle of the intersection of the planetary orbit with the plane of the sky at Proxima's distance, when the Sun--planet distance is increasing.

\begin{figure}
\centering
\includegraphics[width=8.5cm]{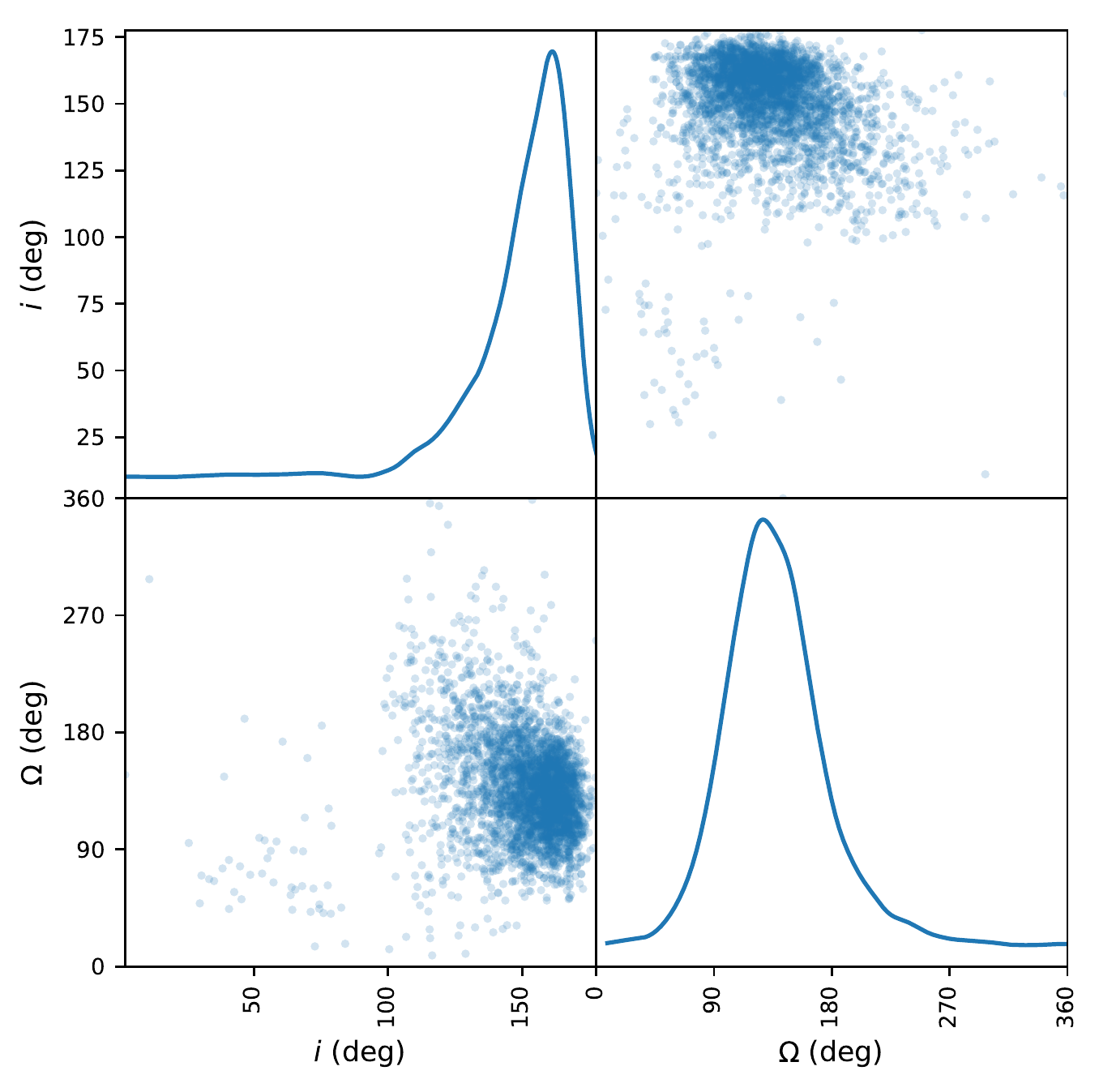}
\caption{Monte Carlo scatter plots (upper right and lower left panels) and kernel density estimates (upper left and lower right panels) of the inclination $i$ and longitude of ascending node $\Omega$ distributions for the prograde orbital solution.
\label{histo-kde}}
\end{figure}

The best-fit orbital parameters and mass of Proxima~c are listed in Table~\ref{Planet-table}, and the MC scatter plots of the distributions of $i$ and $\Omega$ for the prograde solution are shown in Fig.~\ref{histo-kde}.
The inclination of the prograde solution is found to be $\inclP \pm \inclPerr \deg$, corresponding to a mass of $m_c = \Massc^{\Masscerrplus}_{\Masscerrminus}\,M_\oplus$ for Proxima~c, comparable to Uranus and Neptune.
We tested a MC computation without any prior on $i$, for which we obtain a best fit value $i=159\,\deg$ and a planet mass of $16\,M_\odot$, which is highly consistent with the results obtained without the prior.

\begin{figure*}
\centering
\includegraphics[width=14cm]{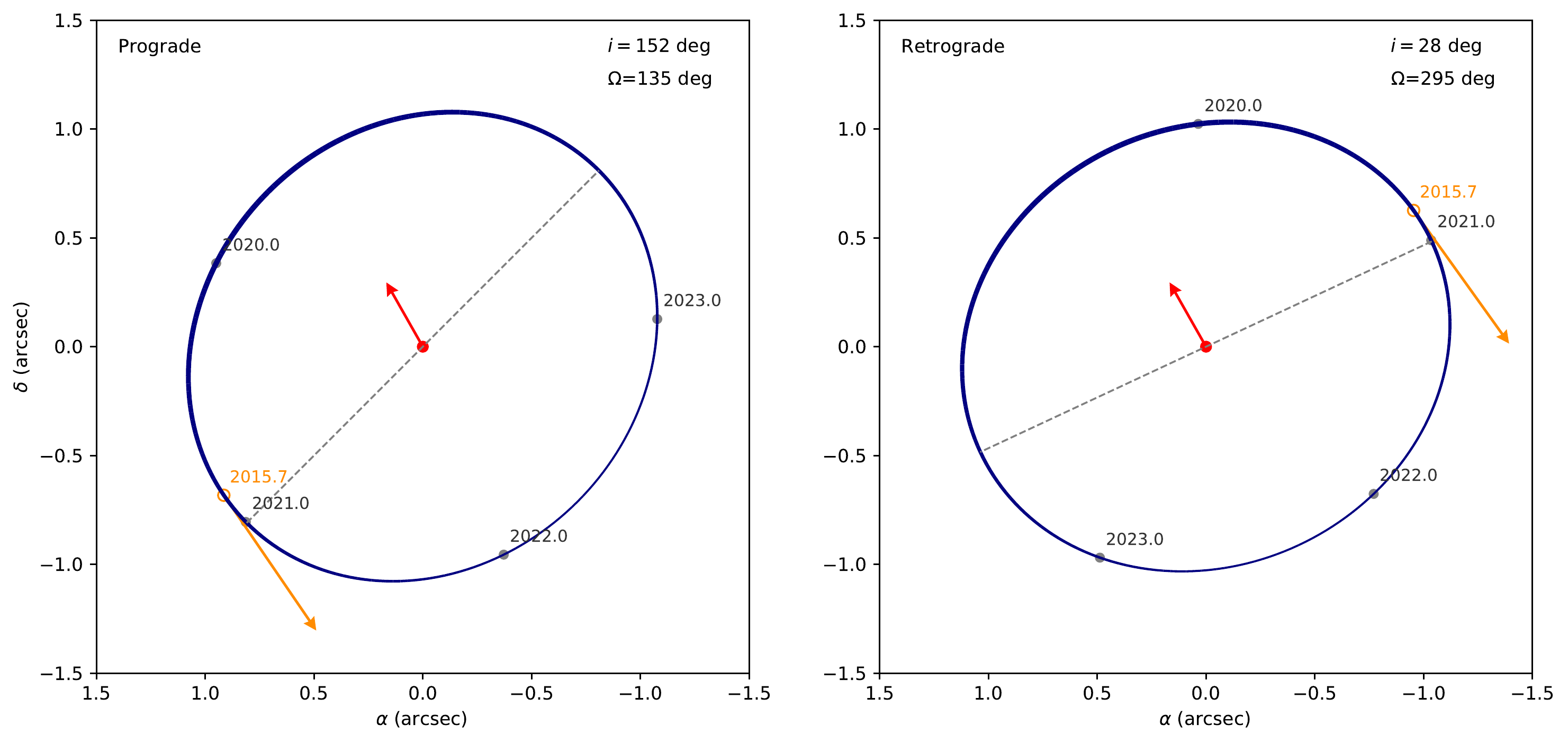}
\caption{Orbital trajectories of Proxima~c corresponding to the best-fit prograde (left panel) and retrograde (right panel) solutions (whose parameters are listed in Table~\ref{Planet-table}).
A thicker line indicates that the planet is closer to the Earth.
The orange arrow shows the velocity vector of Proxima~c at the effective GDR2 epoch, and the red arrow the corresponding reflex velocity of Proxima from its PMa (scaled by $10000\times$).
\label{orbit-plot}}
\end{figure*}

\begin{figure*}
\centering
\includegraphics[width=15cm]{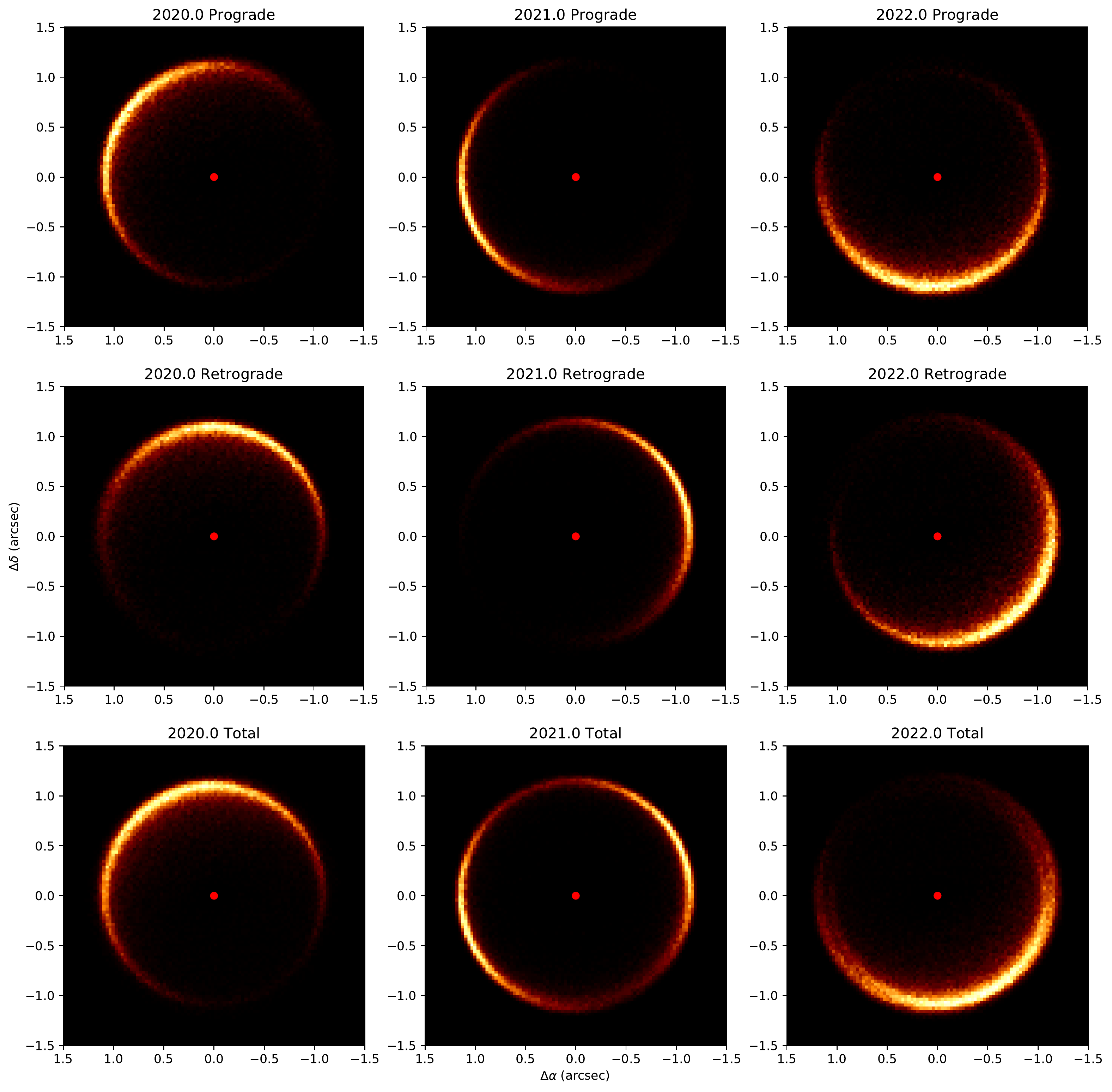}
\caption{Maps of the probability of presence of Proxima~c relative to its parent star for epochs 2020.0, 2021.0, and 2022.0. The maps are presented for the prograde solution (top row), the retrograde solution (middle row), and the cumulated probability of presence with no assumption on the orbital direction (bottom row). The color scale is linear and arbitrarily normalized to the maximum probability in the field at each epoch.
\label{position-plot}}
\end{figure*}

The best-fit prograde and retrograde orbits are displayed in Fig.~\ref{orbit-plot}.
Due to the relatively large uncertainties on $i$ and $\Omega$, this map cannot be used to accurately predict the position of Proxima~c at any time.
However, when the orbital phase of the planet is close to the ascending or descending nodes, its relative position with respect to Proxima is significantly more probable over a relatively narrow arc.
The maps of the probability of presence of Proxima~c for epochs 2020.0, 2021.0 (close to the ascending node), and 2022.0 are shown in Fig.~\ref{position-plot}.

\begin{table}[htp]
\caption{Derived parameters of Proxima c.}
\centering
\begin{tabular}{lcc}
\hline \hline
\noalign{\smallskip}
Quantity & & Value \\
\hline
\noalign{\smallskip}
Orbital radius & $a_c$ & $1.489 \pm 0.049$\,au \\
 & $a_c$ & $1.145 \pm 0.041\,\arcsec$ \\
\noalign{\smallskip}
Minimum mass of planet & $m_c \sin i$  &  $5.7 \pm 1.9\,M_\oplus$ \\
\noalign{\smallskip}
Prograde solution:\\
\ \ \ Orbital inclination & $i$ & $\inclP \pm \inclPerr\,\deg$ \\
\ \ \ Longitude of asc. node & $\Omega$ & $\OmegaPval \pm \OmegaPvalerr\,\deg$ \\
\ \ \ Correlation $(i,\Omega)$ & $\rho(i,\Omega)$ & \correlvalP \\
\noalign{\smallskip}
Retrograde solution:\\
\ \ \ Orbital inclination & $i$ & $\inclR \pm \inclRerr\,\deg$ \\
\ \ \ Longitude of asc. node & $\Omega$ & $\OmegaRval \pm \OmegaRvalerr\,\deg$ \\
\ \ \ Correlation $(i,\Omega)$ & $\rho(i,\Omega)$ & \correlvalR \\
\noalign{\smallskip}
 \noalign{\smallskip}
Mass of planet c & $m_c$ & $\Massc^{\Masscerrplus}_{\Masscerrminus}\,M_\oplus$ \\
 \noalign{\smallskip}
\hline
\end{tabular}
\label{Planet-table}
\end{table}

\section{Discussion\label{discussion}}

In the present analysis, the error budget of the orbital parameters of Proxima c is dominated by the precision of the PMa vector, and more specifically by the GDR2 PM vector of Proxima. The uncertainties on the components of the long-term Hip2-GDR2 PM vector ($\vec{\mu_\mathrm{HG}}$) are approximately four times smaller than those of the short-term PM vector ($\vec{\mu_\mathrm{G2}}$).
 However, the uncertainty on the spectroscopic radial velocity is quite comparable: the mean velocity anomaly of Proxima in the tangential plane over the GDR2 time-span is  $\vec{\Delta \mu} = \left[+1.34 \pm 0.69, +2.37 \pm 1.33\right]$\,m~s$^{-1}$, while the mean radial velocity is $v_r = -0.94 \pm 0.40$\,m~s$^{-1}$. 
The Gaia Early Data Release 3 (EDR3) is expected in the third quarter of 2020. It will bring significant improvement to the precision of the Gaia PM vector, and therefore also the PMa vector, possibly by a factor of more than  approximately two thanks to the longer time base and the decrease in systematic error.
This will provide a comparable improvement to the orbital parameters and mass of Proxima~c.

The inclination of the dust rings identified by \citetads{2017ApJ...850L...6A} ($\approx 45^\circ$) from ALMA observations of Proxima is compatible with our derived inclination.
The position angle of the major axis of the ring ($\approx 140^\circ$) is also in agreement with the position angle of the line of nodes of the orbit of Proxima c.
On a larger scale, we note that the orbit of Proxima in the $\alpha$\,Cen system \citepads{2017A&A...598L...7K} and the orbit of the main components $\alpha$\,Cen A and B are both progrades (counter clockwise), possibly favoring the prograde solution for the orbit of Proxima~c (Table~\ref{Planet-table}).

If we assume the coplanarity of the orbits of the planets Proxima b and c, the de-projected mass of the close-in planet is $m_b = 2.1^{+1.9}_{-0.6}\,M_\oplus$ (adopting $m_b \sin i = 1.0 \pm 0.1\,M_\oplus$ from \citeads{2020SciA....6.7467D}).
It has been suggested that this planet is lying in the habitable zone of Proxima, but this red dwarf is known to experience strong flares \citepads{2018ApJ...855L...2M,2018ApJ...860L..30H}.
\citetads{2019ApJ...884..160V} recently observed repeated, very energetic events using the TESS (Transiting Exoplanet Survey Satellite).
Such high-energy flaring could reduce the chance that Proxima~b hosts life.
However, a high planet mass could help protect its surface from the high-energy radiation and particles, through the preservation of its atmosphere and the possible presence of a magnetic field.
Depending on the greenhouse effect on Proxima~b, the flares could induce adequate temperatures for liquid water that, if the atmosphere is dense enough, would in turn protect its surface from the flares.
We note that \citetads{10.1093/mnrasl/slaa037} suggest that a fraction of the population of microorganisms on Proxima b is able to survive the flares and superflares of Proxima.

\citetads{2019MNRAS.490.5002F} recently presented a combined astrometry and radial velocity analysis for the massive ($3\,M_J$), long-period (45\,years) planet orbiting $\epsilon$\,Ind~A. While the present work does not reach a comparable level of predictive accuracy on the position and mass of the much-less-massive Proxima c, it confirms the  high potential of the combination of ultra high-accuracy astrometry and radial velocity measurements.
As the astrometric signature of orbiting companions is linearly decreasing with distance, emphasis should be placed on radial velocity monitoring of the nearest stars not saturating the Gaia detectors in order to reach the highest possible sensitivity in combination with Gaia astrometry.

\begin{acknowledgements}
The present work has benefited substantially from insightful suggestions by the referee Dr. Timothy D. Brandt, to whom we are grateful.
This work has made use of data from the European Space Agency (ESA) mission {\it Gaia} (\url{http://www.cosmos.esa.int/gaia}), processed by the {\it Gaia} Data Processing and Analysis Consortium (DPAC, \url{http://www.cosmos.esa.int/web/gaia/dpac/consortium}).
This research has made use of Astropy\footnote{Available at \url{http://www.astropy.org/}}, a community-developed core Python package for Astronomy \citepads{2018AJ....156..123A}.
We used the SIMBAD and VizieR databases and catalog access tool at the CDS, Strasbourg (France), and  NASA's Astrophysics Data System Bibliographic Services.
\end{acknowledgements}

\bibliographystyle{aa} 
\bibliography{../biblioAlfCen}

\end{document}